\documentclass[prc,aps,twocolumn,preprintnumbers,amsmath,amssymb,floatfix,superscriptaddress]{revtex4}
\usepackage[usenames]{color}
\usepackage{amsmath}
\usepackage{graphicx}
\usepackage{float,epsfig}
\usepackage{lscape}
\usepackage[english]{babel}
\usepackage{indentfirst}
\usepackage{amsxtra}

\usepackage{supertabular}
\usepackage{multirow}
\usepackage[mathcal]{eucal}  
\usepackage[usenames]{color}
\usepackage{ulem}

\newcommand{\simge}{\hspace*{0.2em}\raisebox{0.5ex}{$>$}
     \hspace{-0.8em}\raisebox{-0.3em}{$\sim$}\hspace*{0.2em}}

\begin{document}

\title{\textbf{Renormalizability of the nuclear many-body problem}\\
with the Skyrme interaction beyond mean field}
\author{C.J. Yang}
\affiliation{Institut de Physique Nucl\'eaire, CNRS/IN2P3, Univ. Paris-Sud, 
Universit\'e Paris-Saclay, 91406 Orsay, France}
\author{M. Grasso}
\affiliation{Institut de Physique Nucl\'eaire, CNRS/IN2P3, Univ. Paris-Sud, 
Universit\'e Paris-Saclay, 91406 Orsay, France}
\author{K. Moghrabi}
\affiliation{Multidisciplinary Physics Lab, Lebanese University, 
Faculty of Sciences I, Hadath, Lebanon}
\affiliation{American University of Science and Technology, Beirut, Lebanon}
\author{U. van Kolck}
\affiliation{Institut de Physique Nucl\'eaire, CNRS/IN2P3, Univ. Paris-Sud, 
Universit\'e Paris-Saclay, 91406 Orsay, France}
\affiliation{Department of Physics, University of Arizona, Tucson, AZ 85721, 
USA}

\begin{abstract}
Phenomenological effective interactions like Skyrme forces are currently
used in mean--field calculations in nuclear physics. Mean--field models 
have strong analogies with the first order of the perturbative many--body
problem and the currently used effective interactions are adjusted at the
mean--field level. In this work, we analyze the renormalizability of the
nuclear many--body problem in the case where the effective Skyrme
interaction is employed in its standard form and the perturbative problem is
solved up to second order. We focus on symmetric nuclear matter and its
equation of state, which can be calculated analytically at this order. It is
shown that only by applying specific density dependence and constraints to
the interaction parameters could renormalizability be guaranteed in principle. 
This indicates that the standard Skyrme interaction does not in general lead 
to a renormalizable theory. For achieving renormalizability, other terms
should be added to the interaction and employed perturbatively only at first
order.
\end{abstract}

\maketitle

\affiliation{Institut de Physique Nucl\'eaire, IN2P3-CNRS, Universit\'e Paris-Sud, 
F-91406 Orsay Cedex, France}

\affiliation{Institut de Physique Nucl\'eaire, IN2P3-CNRS, Universit\'e Paris-Sud, 
F-91406 Orsay Cedex, France}

\affiliation{Institut de Physique Nucl\'eaire, IN2P3-CNRS, Universit\'e Paris-Sud, 
F-91406 Orsay Cedex, France}

\affiliation{Faculty of Sciences, Lebanese University, Beirut, Lebanon} %
\affiliation{American University of Science and Technology, Beirut, Lebanon}

\vskip 0.5cm

\section{Introduction}

Bulk properties of medium--mass and heavy nuclei are very well described by
phenomenological effective interactions treated in the mean--field picture 
\cite{bender}. Despite this success, the necessity of increasing the
accuracy of theoretical predictions in some cases has motivated several
groups to formulate beyond--mean--field models, which explicitly include
more correlations in their formal scheme. Several directions have been
explored, for instance: projection techniques in the framework of the
generator coordinate method \cite{anguiano,robledo,sandulescu,lacroix1};
second random--phase approximation (SRPA) calculations with Skyrme and Gogny
forces \cite{gamba2010,gamba2012}, as well as with an interaction derived
from a realistic force \cite{papak}; particle--vibration coupling (PVC)
techniques with the Skyrme interaction \cite{bernard,colo} and with a
relativistic Lagrangian \cite{litvinova}; multiparticle--multihole
configuration mixing (mpmhCM) methods with both Skyrme \cite{pillet1} and
Gogny \cite{pillet2} forces.

A challenge faced by all these models is how to overcome the overcounting of
correlations when conventional forces or Lagrangians are used. Conventional
forces and Lagrangians are actually designed for mean--field--based models,
and the adjustment of their parameters is performed at this level. Using the
same interactions, with the same values of the parameters, in calculations
where different types of correlations are explicitly taken into account
obviously produces some double counting. In other words, when
beyond--mean--field methods are used, the adjustment of the parameters
should be done at the same level (at the same order) in the perturbative
many--body problem. Otherwise, subtraction procedures should be applied to
cancel the overcounted correlations such as, for instance, the subtraction
method introduced by Tselyaev \cite{tse2007,tse2013} and applied to PVC 
\cite{nonrela1,nonrela2,nonrela3,rela1,rela2,rela3} and SRPA \cite{gamba} 
models. Apart from this general problem, several technical difficulties are
encountered in many of these sophisticated models. Let us mention for
instance the irregularities and the divergences that may be found in
projection calculations \cite{anguiano,bender1,lacroix,duguet} and the
ultraviolet (UV) divergences that are present when zero--range interactions
are employed in SRPA, PVC or mpmhCM calculations (these divergences may be 
eliminated in 
some specific cases by applying the subtraction procedure mentioned above).

The issue of UV divergences in second--order calculations with the
zero--range Skyrme force has been addressed by three of us in the case of
nuclear matter using cutoff-- and dimensional--regularization techniques 
\cite{moghra1,yang1}. New--generation Skyrme--type interactions have been
designed to provide a reasonable equation of state (EOS) for nuclear matter
by including first-- and second--order contributions in the evaluation of
the energy. This approach produces well--defined results that avoid
overcounting.

The specific problem of designing new interactions to be used in
beyond--mean--field calculations can be viewed as a part of a more general
issue: the formulation of an interaction that provides a renormalizable
theory order by order in the perturbative many--body problem.
Renormalizability means that the theory is independent of the 
details of high--energy physics and, in particular, the arbitrary
regularization procedure. High--energy physics eliminated from loops by the
regulator is accounted for in the coefficients of the interactions, which
are then cutoff dependent in such a way as to ensure that observables are
not. Renormalizability is guaranteed once all interactions allowed by the
symmetries of the underlying dynamics are included. The framework to
accomplish this is that of effective field theories (EFTs), which has been
successfully applied to the physics of
light nuclei over the last two decades \cite{EFT1,EFT2}. 

Ensuring renormalizability is, in turn, a step towards an even more general
objective, that of searching for the correct power counting which indicates
the proper hierarchy of allowed interactions. A consistent power counting
generates at each order enough interactions so that any remaining
regularization dependence can be eliminated with a sufficiently high value
for the regulator parameter. Thus, imposing renormalizability is a guide for
theory construction, the best-known example being the development of the
electroweak theory known as the Standard Model. A nuclear example is provided 
by Pionless EFT, where the existence of a three--body force in leading order 
was discovered by demanding renormalizability of the theory's description of 
the three-body system \cite{threebozos1,threebozos2,threebozos3}. 
Unfortunately, the renormalization of Chiral EFT, which extends Pionless EFT
to momenta comparable to the pion mass, is not fully understood even in
few-nucleon systems \cite{pionfulrenorm1,pionfulrenorm2,yang-rev}.

The successes of mean--field models suggest that there should be a
controlled expansion around it. However, renormalizability has not yet been
extensively explored in the case of phenomenological effective interactions
like Gogny and Skyrme forces. In this exploratory study, we focus on the
zero--range Skyrme force, which bears formal similarities with the
interactions in Pionless EFT. We are thus implicitly assuming that
non--relativistic nucleons are the relevant degrees of freedom for the
low--energy dynamics of the nuclei of interest. The analysis is performed by
including first-- and second--order contributions in the EOS of symmetric
nuclear matter. The objective is to reveal the implications of demanding
renormalizability through a redefinition of the existing parameters at
second order. A similar procedure can be followed for more complex forces,
higher orders, different isospin asymmetries, and finite nuclei.

\section{Renormalization}

We consider the standard Skyrme force \cite{vauthe}, which contains central,
density--dependent, spin--orbit, and velocity--dependent terms of zero
range. The spin--orbit term does not contribute in infinite matter at first
order, but in general does provide a second--order contribution to the EOS 
\cite{spinorbit}. The contribution of this term under dimensional
regularization can be found in Ref. \cite{Kaiser}. For the sake of
simplicity, in this first exploratory study of the renormalizability of the
problem, we have omitted this term in the interaction as also done in Ref. 
\cite{yang1}. 
In contrast, we keep the density--dependent part of the Skyrme 
interaction, even though 
such a term might be problematic 
in connection to the so--called self--interaction problem. 
It was recognized already in the 70s 
\cite{stringari} 
that only density--independent contact forces 
allow one to satisfy specific antisymmetry conditions 
in the solution of random--phase--approximation equations. 
The violation of such conditions is associated with a violation of the 
Pauli principle generated by spurious contributions coming from the 
interaction of a particle with itself.
This self--interaction problem was discussed again more recently 
\cite{bender,lacroix,bender1,duguet,erler,drut,sadoudi,chamel}. 
For instance, a strategy to solve pathologies produced by the 
self--interaction problem was suggested in Ref. \cite{lacroix} for 
the beyond--mean--field case of the generator--coordinate method. 
We retain the density--dependent term 
because it is known to be necessary to describe well the equilibrium point of 
symmetric matter not only in first order 
but also in second order, as 
discussed in Refs. \cite{yang1,Kaiser}. 
Alternative terms such as a real three--body force would 
yield much more involved calculations. 
It is also worth mentioning that, by including the rearrangement terms 
associated with the density--dependent force in the computation of the 
second--order EOS, as we do here, the Hugenholtz-Van Hove theorem \cite{huge} 
is satisfied.

We define the incoming and outgoing relative momenta, 
$\vec{k}=(\vec{k}_{1}-\vec{k}_{2})/2$ and 
$\vec{k}^{\prime }=(\vec{k}_{1}^{\prime }-\vec{k}_{2}^{\prime})/2$,
where $\vec{k}_{i}^{(\prime )}$ denotes the momentum of nucleon 
$i^{(\prime )}$. 
We also introduce the spin--exchange operator 
$P_{\sigma}=(1+\vec{\sigma}_{1}\cdot \vec{\sigma}_{2})/2$ in terms of the 
spin $\vec{\sigma}_{i}/2$ of nucleon $i$. 
We deal only with symmetric nuclear matter,
for which the density $\rho $ and the Fermi momentum $k_{F}$ (the same for
neutrons and protons) are related by the relation 
$k_{F}=(3\pi^{2}\rho/2)^{1/3}$.
In terms of these quantities,
the interaction is written as 
\begin{widetext}
\begin{eqnarray}
V(\vec{k},\vec{k}^{\prime})&=& t_0(1+x_0P_{\sigma})
+T_3 (1+x_3P_{\sigma}) k_F^{3\alpha}
+ \frac{t_1}{2} (1+x_1P_{\sigma}) 
\left(\vec{k}'^{2}+\vec{k}^{2}\right)
+ t_2(1+x_2P_{\sigma})\vec{k}'\cdot \vec{k}, 
\label{inte}
\end{eqnarray}
\end{widetext}
where $\alpha $ is a real number. The usual Skyrme parameters $t_{0,1,2}$
and $x_{0,1,2,3}$ are present, while the parameter $T_{3}$ is defined in
terms of the Skyrme parameter $t_{3}$ as 
$T_{3}=(2/(3\pi ^{2}))^{\alpha}t_{3}/6$. 
The $T_{3}$ term describes the so--called density--dependent
part of the interaction, which is necessary to ensure the correct
description of the saturation point and of the compressibility modulus of
symmetric matter not only at the mean--field level but also at second order 
\cite{Kaiser,yang1}.

Our regulator is chosen, as in Ref. \cite{yang1}, as a cutoff $\lambda $ put
on the outgoing relative momentum $\vec{k}^{\prime }$ , $\lambda=\tilde{%
\lambda}/k_F$. Other regulators generate
terms of the same form but with different coefficients. Dimensional
regularization with standard subtraction procedures sets several of these
coefficients to zero and tends to hide a potential lack of
renormalizability, one example being the two-body system with resonant $p$%
-wave interactions \cite{halo1, halo2}. For this reason, we do not employ 
such type of renormalization here.

\begin{figure}
\includegraphics[scale=0.5]{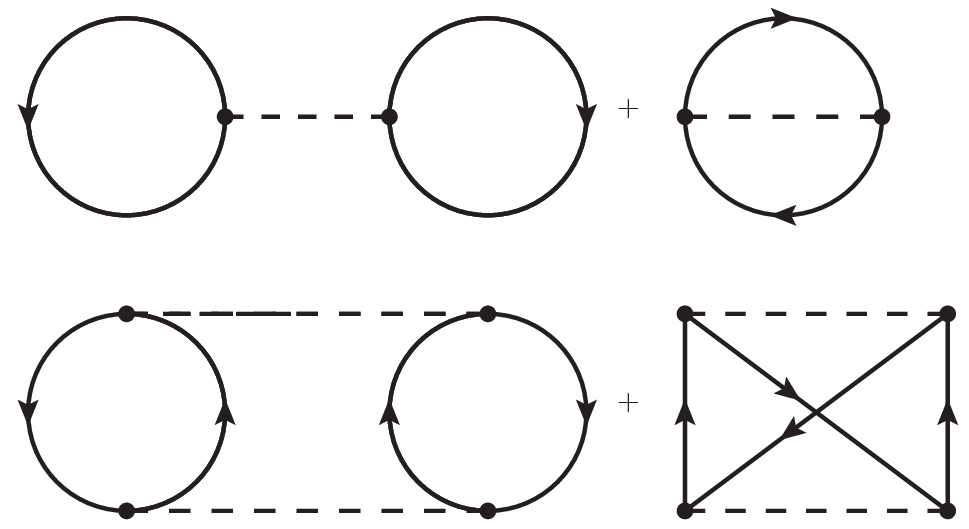}
\caption{First--order (upper line) and second--order (lower line) 
contributions to the energy of nuclear matter. 
Direct (exchange) terms are shown on the left (right).
Particles and holes are denoted by oriented solid lines
and the interaction by a dashed line.}
\label{diagrams}
\end{figure}

The EOS for symmetric matter is given, up to second order, by 
the diagrams shown in Fig. \ref{diagrams}. 
The upper (lower) line displays first-- (second--) order diagrams,
while direct (exchange) contributions are shown on the left (right) column.
The evaluation of these diagrams gives for the energy per nucleon
\begin{eqnarray}
\frac{E}{A}(k_{F},\tilde{\lambda}) &=&\frac{3\hbar ^{2}}{10m}k_{F}^{2}+\frac{%
t_{0}}{4\pi ^{2}}k_{F}^{3}+\frac{T_{3}}{4\pi ^{2}}k_{F}^{3+3\alpha }  \notag
\\
&&+\frac{\theta _{s}}{4\pi ^{2}}k_{F}^{5}+\frac{\Delta E^{(2)}}{A}(k_{F},%
\tilde{\lambda}).  
\label{eos}
\end{eqnarray}%
The first term of Eq. (\ref{eos}) is the
kinetic contribution ($m$ is the nucleon mass) and the following three terms
are first--order, with 
\begin{equation}
\theta _{s}=\frac{1}{10}\left[ 3t_{1}+t_{2}(5+4x_{2})\right] .
\end{equation}
The last term of Eq. (\ref{eos}) collects the second--order contributions,
which depend on the momentum cutoff. The expression for 
$\Delta E^{(2)}(k_{F},\tilde{\lambda})/A$ in symmetric matter can be found in 
Ref. \cite{yang1}, 
including the contributions coming from rearrangement terms 
in the prescription of Ref. \cite{carlsson}. 
The asymptotic behavior ($\tilde{\lambda}\gg k_{F}$), which has a polynomial 
form in the cutoff, is also given. We have checked that this 
asymptotic polynomial form practically coincides with the full expressions
starting from $\tilde{\lambda}\simge$ 1 fm$^{-1}$. The
asymptotic expression can be split into three terms, 
\begin{eqnarray}
\frac{\Delta E^{(2)}(k_{F},\tilde{\lambda})}{A} &=&\frac{\Delta
E_{f}^{(2)}(k_{F})}{A}+\frac{\Delta E_{a}^{(2)}(k_{F},\tilde{\lambda})}{A} 
\notag \\
&&+\frac{\Delta E_{d}^{(2)}(k_{F},\tilde{\lambda})}{A},  \label{sum}
\end{eqnarray}%
where the subscripts $f$, $a$, and $d$ stand for \textquotedblleft
finite\textquotedblright , \textquotedblleft absorbed\textquotedblright ,
and \textquotedblleft divergent\textquotedblright\ and denote, respectively,
the finite part, the contribution where the cutoff dependence can be
absorbed with a redefinition of the interaction parameters, and the part
that cannot in general be regrouped with mean--field terms and thus diverges
when $\tilde{\lambda}\rightarrow \infty$. 

Denoting by $m^{\ast }$ the
nucleon effective mass in symmetric matter and using its mean--field 
expression, as done in Ref. \cite{yang1}, 
\begin{equation}
m^{\ast }=m\left(1+\frac{5m}{6\pi^2\hbar ^{2}}k_F^3\theta_s\right)^{-1},
\label{ms}
\end{equation}
the three contributions in Eq. (\ref{sum}) can be written as 
\begin{widetext}
\begin{equation}
\frac{\Delta E_{f}^{(2)}(k_{F})}{A}=\frac{3m^{\ast }}{2\pi ^{4}\hbar ^{2}}
k_{F}^{4}\left[ A_{0}+A_{1}T_{3}k_{F}^{3\alpha}+A_{2}T_{3}^{2}k_{F}^{6\alpha }
+A_{3}k_{F}^{2}+A_{4}T_{3}k_{F}^{2+3\alpha}+A_{5}k_{F}^{4}\right],
\label{terms1}
\end{equation}
\begin{equation}
\frac{\Delta E_{a}^{(2)}(k_{F}, \tilde{\lambda} )}{A}=
-\frac{m}{8\pi^{4}\hbar ^{2}}\tilde{\lambda} k_{F}^{3}\left[ B_{0}(\tilde{\lambda})
+B_{1}(\tilde{\lambda} )T_{3}k_{F}^{3\alpha }+B_{2}(\tilde{\lambda})k_{F}^{2}\right],
\label{terms2}
\end{equation}
and
\begin{equation}
\frac{\Delta E_{d}^{(2)}(k_{F},\tilde{\lambda} )}{A}=
-\frac{m^{\ast }}{8\pi^{4}\hbar ^{2}}\tilde{\lambda} k_{F}^{3}
\left[ C_{0}T_{3}^{2}k_{F}^{6\alpha}+C_{1}T_{3}k_{F}^{2+3\alpha }+C_{2}k_{F}^{4}\right]
+\frac{m^{\ast }-m}{m}
\frac{\Delta E_{a}^{(2)}(k_{F}, \tilde{\lambda} )}{A}.
\label{terms3}
\end{equation}
\end{widetext} 
The coefficients $A_{i}$, $B_{i}(\tilde{\lambda})$, and $C_{i}$ are
combinations of Skyrme parameters and (for the coefficients $B_{i}$)
cutoff, which are shown explicitly in the Appendix.

From the $k_{F}$ dependence in Eq. (\ref{terms2}), one
sees that $\Delta E_{a}^{(2)}(k_{F},\tilde{\lambda} )/A$ may be regrouped 
with mean--field terms in the EOS. 
The cutoff dependence can be absorbed in the bare interaction parameters 
in the form of 
renormalized parameters (denoted by superscript $^{R}$)
\begin{equation}
t_{0}^{R}=t_{0}(\tilde{\lambda})
-\frac{m\tilde{\lambda} }{2\pi^{2}\hbar ^{2}}B_{0}(\tilde{\lambda}),  
\label{coeff1}
\end{equation}
\begin{equation}
T_{3}^{R} =T_{3}(\tilde{\lambda})
\left[1-\frac{m\tilde{\lambda}}{2\pi^{2}\hbar^{2}}B_{1}(\tilde{\lambda})\right],
\label{coeff2}
\end{equation}
and
\begin{equation}
\theta _{s}^{R} =\theta _{s}(\tilde{\lambda})
-\frac{m\tilde{\lambda}}{2\pi ^{2}\hbar ^{2}}B_{2}(\tilde{\lambda}).  
\label{coeff3}
\end{equation}
Assuming that the expansion beyond mean field is well defined,
one can directly replace the bare parameters in 
Eqs. (\ref{terms1}) and (\ref{terms3})
by the renormalized ones. We indicate this by a superscript $^{R}$ in the
coefficients $A$ and $C$. 
With this replacement, we induce a change 
of the original equations, but 
such modifications are of higher order (at least third--order induced terms) 
and may thus be neglected if
only terms up to second order are retained in the EOS, as we require here.

Equation (\ref{terms3}) shows that the cutoff dependence in 
$\Delta E_{d}^{(2)}(k_{F},\tilde{\lambda} )/A$ cannot be similarly handled. 
Adoption of the mean--field effective mass \eqref{ms} brings 
a $k_F$ dependence in the denominator 
that cannot be absorbed in density--independent parameters.
This forces us to set 
\begin{equation}
\theta_s^R=0
\label{thetaR=0}
\end{equation}
to eliminate the last term in Eq. (\ref{terms3}).
It also follows that $m^{\ast }=m$ throughout Eqs. (\ref{terms1}) and 
(\ref{terms3}). 
The remaining contributions in Eq. (\ref{terms3}) must now be handled 
by imposing renormalizability. 
The most dangerous dependence is $\propto k_F^7$, which cannot be modified by
a choice of $\alpha$ and has a positive 
coefficient, $C_2\ge 0$. Because $C_0> 0$, the $k_F^{3+6\alpha}$ dependence
is also constrained.

The first possibility is that the divergent terms cancel among themselves.
This requires
$\alpha=2/3$, in which case all three divergent terms have the same 
$k_F$ dependence. Then, Eq. (\ref{terms3}) vanishes if and only if
\begin{equation}
 C_{0}^R(T_{3}^{R})^2+C_{1}^R T_{3}^R +C_{2}^R=0.  
\label{cancel}
\end{equation}
However, we found that, even without the constraint $\theta_s^R=0$, the 
discriminant of the above equation is always less or equal to 0. 
Thus, no real set of parameters satisfies Eq. (\ref{cancel}), 
except for the trivial case $t_1^R=t_2^R=T_3^R=0$. 
In this case, the only divergence is absorbed through Eq. (\ref{coeff1}), 
and the resulting EOS is finite. The
renormalized EOS for symmetric nuclear matter evaluated up to second order
is given by 
\begin{equation}
\frac{E}{A}(k_{F}) = 
\frac{3\hbar ^{2}}{10m}k_{F}^{2}+\frac{t_{0}^{R}}{4\pi ^{2}}k_{F}^{3}  
+\frac{3m}{2\pi^{4}\hbar^{2}} A_0^R k_F^4. 
\label{eafull1}
\end{equation}
As shown in the Appendix, $A_0^R$ contains $x_0$, which in symmetric matter does
not contribute in first order.
We can take $A_0^R\geq 0.069 (t_0^{R})^2$ as an independent
parameter so that Eq. \eqref{eafull1} has only two parameters to adjust,
if we interpret $m$ as the in-vacuum nucleon mass.
Unless $m$ is taken as a free, negative parameter,
this ``$t_0$ model'' does not lead to
saturation at mean--field level, meaning 
that the saturation point requires a second--order contribution
comparable to first order. 

For a more meaningful model, not all divergent terms vanish,
and thus must be absorbed in mean--field terms. 
The only possibility for the $k_F^7$ divergence is the $T_3$ term
(in which case $\alpha=4/3$), but then the $k_F^{3+6\alpha}$
divergence cannot be eliminated. Thus we must consider
the ``$t_0-t_3$ model'' where $t_1^R=t_2^R=0$ and
\begin{equation}
C_2 ^R=0, \qquad C_1^R=0.
\label{cancel2}
\end{equation}
This in turn implies $B_2(\tilde\lambda)=0$, and 
$A_3^R=A_4^R=A_5^R=0$.
In this model, saturation is obtained at first order thanks
to the $T_3$ term.

Now, only for specific values of $\alpha\ne 0$ can we eliminate the
$k_{F}^{3+6\alpha}$ divergent term.

\begin{enumerate}

\item If $\alpha =-1/6$ the $k_{F}^{3+6\alpha}$ divergent term can be absorbed
into a renormalized--mass term by a choice of the bare mass, 
that is,
\begin{equation}
\qquad\frac{1}{m^{R}}=\frac{1}{m(\tilde\lambda)}
\left\{ 1-\frac{5\tilde{\lambda}m^2(\tilde\lambda)}{12\pi^{4}\hbar^{4}} 
(T_{3}^{R})^2\left[\left(\frac{175}{192}\right)^2+x_3^2\right]\right\}.
\label{mr}
\end{equation}
Since we keep only terms up to second order in the EOS, 
we can directly replace $m$ by $m^{R}$ in Eqs. (\ref{terms1})
and (\ref{terms2}) and neglect the induced higher--order contributions.
The finite $A_2$ term in Eq. \eqref{terms1} is now $\propto k_F^3$
and can also be absorbed in $t_{0}^{R}$, that is, we modify
Eq. \eqref{coeff1} to
\begin{equation}
t_{0}^{R}=t_{0}(\tilde{\lambda})
-\frac{m\tilde{\lambda} }{2\pi^{2}\hbar ^{2}}B_{0}(\tilde{\lambda})  
+ \frac{6m^{R}}{\pi^{2}\hbar ^{2}}A_2^R (T_3^{R})^2.
\label{coeff1prime}
\end{equation}
The EOS becomes
\begin{widetext}
\begin{equation}
\frac{E}{A}(k_{F}) =\frac{3\hbar^{2}}{10m^{R}}k_{F}^{2}
+\frac{T_{3}^{R}}{4\pi ^{2}}k_{F}^{5/2}  
+\frac{t_{0}^{R}}{4\pi ^{2}}k_{F}^{3}
+\frac{3m^{R}}{2\pi ^{4}\hbar ^{2}}
\left(A_1^R T_3^R k_{F}^{7/2}+ A_0^R k_{F}^{4}\right).
\label{eafull16}
\end{equation}
\end{widetext} 
The pure second--order terms $A_1^R$ and $A_0^R$ are proportional
to $t_{0}^{R}$ and $(t_{0}^{R})^2$, respectively, but they also depend
on the spin coefficients $x_{0,3}$.
In symmetric nuclear matter, where $x_{0,3}$ do not appear in first order,
we can treat $A_1^R$ and $A_0^R\geq 0.069 (t_0^{R})^2$ as independent parameters.

\item If $\alpha =1/3$ the $k_{F}^{3+6\alpha}$ divergent term 
and the finite $A_1$ term in Eq. \eqref{terms1} 
can be absorbed into $\theta_s^R$,
changing Eq. \eqref{coeff3} into
\begin{equation}
\theta_{s}^{R} =\theta_{s}(\tilde{\lambda})
-\frac{m\tilde{\lambda}}{2\pi ^{2}\hbar^{2}} C_{0}^R (T_{3}^{R})^2
+\frac{6m}{\pi ^{2}\hbar^{2}} A_1^R T_3^R.
\label{coeff3prime}
\end{equation}
This is a fine-tuned scenario where at least one
of the bare $t_1(\tilde{\lambda})$ and $t_2(\tilde{\lambda})$ parameters 
is not zero, but their renormalized values are.
Although unusual, it is a scenario 
similar to Pionless EFT at the unitarity limit,
where the bare coefficient of the non-derivative two-body contact interaction
absorbs a linear divergence but its inverse renormalized value is zero
\cite{Konig:2016utl}. 
Now, the finite $A_0$ term in Eq. \eqref{terms1} 
is $\propto k_F^4$ and can be absorbed in $T_{3}^{R}$ 
by replacing Eq. \eqref{coeff2} with
\begin{equation}
T_{3}^{R} =T_{3}(\tilde{\lambda})
\left[1-\frac{m\tilde{\lambda}}{2\pi^{2}\hbar^{2}}B_{1}(\tilde{\lambda})\right]
+\frac{6m}{\pi ^{2}\hbar ^{2}}A_0^R.
\label{coeff2prime}
\end{equation}
The EOS is in this case
\begin{widetext}
\begin{equation}
\frac{E}{A}(k_{F}) =\frac{3\hbar^{2}}{10m}k_{F}^{2}
+\frac{t_{0}^{R}}{4\pi ^{2}}k_{F}^{3}
+\frac{T_{3}^{R}}{4\pi ^{2}}k_{F}^{4}  
+\frac{3m}{2\pi ^{4}\hbar ^{2}}A_2^R (T_3^{R})^2k_{F}^{6}.
\label{eafull13}
\end{equation}
\end{widetext} 
Again, here we can take $A_2^R>0$ as an independent parameter.

\end{enumerate}

Note that for the $t_0-t_3$ models we chose to absorb finite terms in
Eqs. \eqref{coeff1prime} and \eqref{coeff2prime}, just as
in Eq. \eqref{coeff3prime}.
This is done consistently with the underlying assumption
that there is an expansion around the mean field.
As a consequence, the $k_F^3$ and $k_F^4$ terms in 
Eqs. \eqref{eafull16} and \eqref{eafull13},
respectively, are unconstrained.
Had we not absorbed these terms, there would be additional
pieces $\propto k_F^3$ and $\propto k_F^4$ with a fixed
dependence on, respectively, $(T_3^{R})^2$ and $(t_0^{R})^2$, which further
constrain the EOS. The difference between absorbing these finite terms
or not provides an estimate of the error stemming from 
our assumption of convergence around the mean field.

There is no other possibility to eliminate the divergent terms
while retaining an expansion where second order does not overcome first
order. As pointed out above,
assuming higher orders provide smaller contributions,
the coefficients of the second--order terms can be replaced by 
their finite, renormalized values. One cannot,
for example, cancel the divergent terms against other second--order 
contributions.

\section{Fits}

Many Skyrme parametrizations exist.
Typical values of the parameter $\alpha$ range from 1/6,
for instance in the Saclay-Lyon forces \cite{sly1,sly2}, up to 1, 
for instance in SIII \cite{siii}. 
Not all of the three values of $\alpha$ that provide a renormalizable force
fall in this range, but we do not discard $\alpha=-1/6$ immediately.
Instead we judge the phenomenological promise of the three
possibilities by fitting a successful EOS.
Because of the overcounting 
problem raised earlier, 
it is not expected that Eqs. (\ref{eafull1}), (\ref{eafull16}),
and (\ref{eafull13})
will lead to a reasonable EOS when one uses 
parameters extracted at mean--field level. 
However, if the expansion converges,
changes in the renormalized parameters (but not the bare parameters,
which depend on the arbitrary cutoff) should be relatively small.

We have performed $\chi^2$ fits of our EsOS
by choosing the SLy5 \cite{sly1,sly2}
mean--field EOS as a benchmark.
We fit in each case $N=18$ energies $E_{i}$, in the range of densities
between 0 and 0.3 fm$^{-3}$, to the SLy5 mean--field reference points 
$E_{i,ref}$, with 
\begin{equation}
\chi ^{2}=\frac{1}{N-1}\sum_{i=1}^{N}
\left(\frac{E_{i}-E_{i,ref}}{0.01\times E_{i,ref}}\right)^{2}.  
\label{chi2}
\end{equation}
The fitted parameters and the associated $\chi^{2}$ values are listed in
Tables \ref{fit_sly51}, \ref{fit_sly53}, and \ref{fit_ea2}. 
In the case $\alpha=-1/6$, a renormalized mass enters in the EOS and we have 
treated it as a free parameter to adjust, as we do for other renormalized 
parameters. 
For the other cases, we first perform the fit with $m=939$ MeV, and 
if a satisfactory result cannot be obtained, we turn the mass 
into a free parameter.
For the cases $\alpha=2/3$ and $\alpha=1/3$ only the magnitudes
of $x_{0,3}^R$ can be determined.

\begin{center}
\begin{table}[tbp]
\centering
\begin{tabular}{c c c c}
\hline\hline
$m$  &  $t_0^R$  & $|x_0^R|$ & $\chi^2$ \\ 
(MeV) &  (MeV fm$^{3}$) &  &  \\
\hline
939 &  $-358.16$ & $<10^{-4}$ & 346850 \\ 
$-969.55$ & 212.28 & $<10^{-4}$ & 15989 \\ 
 \hline\hline
\end{tabular}%
\caption{Parameter sets obtained by fitting the renormalized second--order
EOS for the case $t_{1}^R=t_{2}^R=T_{3}^R=0$ with $\alpha=2/3$, 
Eq. (\protect\ref{eafull1}),
to the SLy5 mean--field EOS. 
The upper line refers to the fit where the mass is taken
at its in-vacuum value, while for the lower line 
the mass is adjusted. }
\label{fit_sly51}
\end{table}
\end{center}

\begin{center}
\begin{table}[tbp]
\centering
\begin{tabular}{cccccc}
\hline\hline
$m^R$  & $t_0^R$ & $T_3^R$  &$x_0^R$ &$x_3^R$ & $\chi^2$   \\ 
(MeV) & (MeV fm$^{3}$) &  (MeV fm$^{5/2}$) &  & & \\
\hline 
$591.9$ & $793.15$  & $-1570.8$ & $1.465$ & $-0.1759$ &  $<0.1$\\
\hline\hline
\end{tabular}
\caption{Parameter set obtained by fitting 
the renormalized second--order EOS
for the case $t_{1}^R=t_{2}^R=0$ with $\alpha=-1/6$, Eq. (\ref{eafull16}),
to the SLy5 mean--field EOS. The renormalized mass is treated as a 
free parameter.}
\label{fit_sly53}
\end{table}
\end{center}

\begin{table}[tbp]
\centering
\begin{tabular}{ccccc}
\hline\hline
$m$  & $t_0^R$  & $T_3^R$ &$|x^R_3|$   & $\chi^2$   \\
(MeV) & (MeV fm$^{3}$) & (MeV fm$^{4}$) &   & \\
\hline
939  & $-1244.1$  & $247.11$ & $<10^{-4}$ & $1364$ \\
\hline
23845  & $-580.16$  & $46.248$ & $<10^{-2}$ & $188$ \\
\hline\hline
\end{tabular}
\caption{Parameter sets obtained by fitting
the renormalized second--order EOS for the case 
$t_{1}^R=t_{2}^R=0$ with $\alpha=1/3$, Eq. (\ref{eafull13}), 
to the SLy5 mean--field EOS.  
The upper line refers to the fit where the mass is taken
at its in-vacuum value, while for the lower line 
the mass is adjusted.}
\label{fit_ea2}
\end{table}

As one can see from Table \ref{fit_sly51}, the fits obtained for the case 
$t_{1}^R=t_{2}^R=T_{3}^R=0$ with $\alpha=2/3$
are totally unsuccessful. 
The magnitude of $x_0^R$ is found to be extremely small.
Adjusting the mass produces a smaller
$\chi^2$, which is however still very large.
More alarming, the mass becomes negative.
The failure of this fit was anticipated by the fact that
for $m>0$
even the saturation point requires large second--order contributions.

\begin{figure}
\includegraphics[scale=0.3]{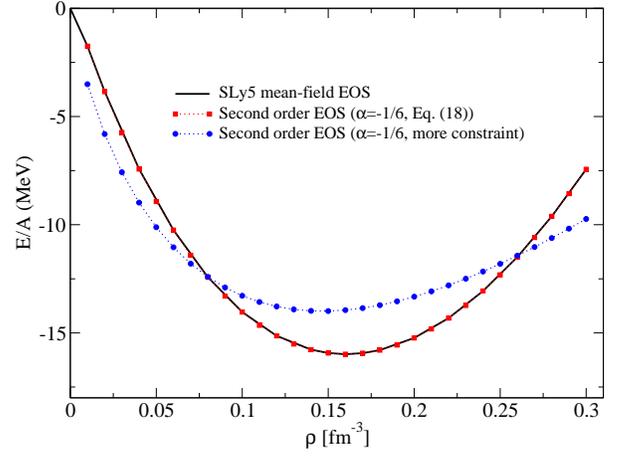}
\caption{EOS for the case $t_{1}^R=t_{2}^R=0$ with $\alpha=-1/6$. 
The fitted second--order EOS Eq. (\ref{eafull16}) (red squares)
is compared to the benchmark EOS (black solid line).
Also shown (blue circles) is a more constrained fit,
where $t_0^R$ is not redefined according to Eq. \eqref{coeff1prime},
which gives an estimate of the potential size of higher--order effects.}
\label{fitfinalm}
\end{figure}

In contrast, for $t_{1}^R=t_{2}^R=0$ with $\alpha=-1/6$, the fit is excellent,
as can be seen also from Fig. \ref{fitfinalm},
where the fit outcome is compared to the SLy5 curve.
This fit is interesting for several reasons.
First, the renormalized mass is somewhat smaller than in-vacuum,
but not so much so that it necessarily invalidates the
non--relativistic approximation.
Second, this value of $\alpha$ inverts the roles of $t_0^R$ and $T_3^R$
in first--order saturation, which now requires $T_3^R<0$ and
$t_0^R>0$. The second--order fit is qualitative stable
in the sense that these signs do not change.
Third, the magnitudes, $|T_3^R|\sim (50 \, \mathrm{MeV})^{-3/2}$
and $t_0^R\sim (100 \, \mathrm{MeV})^{-2}$, have very natural
sizes.
Fourth, the values of $x_{0,3}$ are not particularly small or large,
just as in the usual mean--field EsOS.
Despite these interesting features, second--order effects are significant.
To see this, we have also performed a fit
with the extra constraint that results from not absorbing
the finite term in Eq. \eqref{coeff1prime}. 
In this case the fit deteriorates significantly, as 
shown in Fig. \ref{fitfinalm}.
The $\chi^{2}=2319$ is large,
with opposite signs for $m^R$, $T_3^R$, and $t_0^R$
and negligible $x_{0,3}$.
Although this fit requires a large change in parameters when
going from first to second order, 
the difference between the two fits should be considered as 
a conservative estimate of the error band from higher--order effects.
Clearly, the goodness of the fit based on Eq. (\ref{eafull16}) is not 
representative of what can be obtained in a systematic expansion 
of this model around the mean field.

\begin{figure}
\includegraphics[scale=0.3]{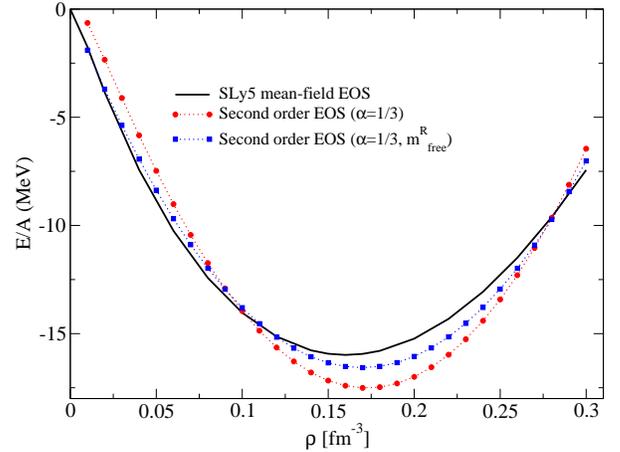}
\caption{EOS 
for the case $t_{1}^R=t_{2}^R=0$ with $\alpha=1/3$. 
The fitted second--order EOS, Eq. (\ref{eafull13}), 
with $m=939$ MeV (red circles) and with $m$ as a free parameter (blue squares) 
are compared to the benchmark EOS (black solid line).}
\label{fitalpha=1/3}
\end{figure}

The remaining case, $t_{1}^R=t_{2}^R=0$ with $\alpha=1/3$,
produces fits of intermediate quality,
shown in Fig. \ref{fitalpha=1/3}.
With the in--vacuum mass, the best-fit EOS is qualitatively
correct, but overbinds with too-large saturation density.
The $\chi^2$ is reduced considerably by allowing $m$ to increase
to a large value, but the fit still overbinds.
The signs of $t_0^R$ and $T_3^R$ are the ones needed for saturation 
in first order, but $|x_3|$ is very small.
When we perform a more constrained fit without absorbing a
finite term in Eq. \eqref{coeff2prime}, the $\chi^2$ soars to
$\chi^2>6000$, again indicating large changes with order.

\section{Conclusion}

It is interesting that we were able to find renormalized
EsOS in second order based on Skyrme forces with $\alpha=-1/6$
and $\alpha=1/3$,
which could potentially serve for a description of finite nuclei.
Of course, before a claim of phenomenological
success can be made, other nuclear--matter properties
(such as the neutron--matter EOS
and the density-dependent symmetry energy) need to be investigated.
Moreover, in both cases there are indications that
higher--order effects will be significant.
Both are specific scenarios in the simple $t_0-t_3$ model,
but with unexpected renormalization features.
For $\alpha=1/3$, 
the renormalization requirements \eqref{cancel2} and \eqref{coeff3prime}
imply
that the derivative interactions at first order decrease
as the cutoff increases, representing a fine tuning
in the $S$--wave energy corrections and in the $P$--wave interaction.
For $\alpha=-1/6$, the mass has to be renormalized
according to Eq. \eqref{mr}, a standard occurrence in
quantum field theory but not in approaches to nuclear matter.
It is not a coincidence that the renormalization
requirements involve the singular two-derivative
two--body terms and the term in Eq. \eqref{inte}
that depends explicitly on the density and is ascribed to few--body forces.
At least one of these terms is required for saturation
and, it is believed, both are needed for a good fit 
at the mean--field level.
It is not at all obvious
that the requirement of renormalization can be fulfilled at higher
orders with such a constrained set of interactions.

A more general renormalization would be achieved only if all the 
cutoff-dependent second--order terms could be regrouped with 
first--order terms without extra constraints. 
For this, additional terms should be added to the interaction. 
From the $k_F$ dependence of the terms in Eq. (\ref{terms3}), we can 
recognize which terms
should be added to the interaction to provide the same $k_F$ dependence in
the EOS. For example, in the case $\alpha=1$ 
(which at mean--field level is a proxy for the three--body force),
for the $k_F^7$ terms one would need a two--body term
of the type $\vec{\nabla}^4 \delta(\vec{r}_1-\vec{r}_2)$; 
for the $k_F^8$ dependence, a three--body term of the type 
$\vec{\nabla}^2 \delta(\vec{r}_1-\vec{r}_2)\delta(\vec{r}_2-\vec{r}_3)$; 
and, finally, for the $k_F^9$ dependence, a four--body term 
$\delta(\vec{r}_1-\vec{r}_2)\delta(\vec{r}_2-\vec{r}_3) 
\delta(\vec{r}_1-\vec{r}_4)$. 

The inclusion of such additional terms would provide of course a much more
complicated interaction and calculations would become more difficult to 
perform in practice. More importantly, if these additional terms are
treated on the same footing as the terms in Eq. (\ref{inte}), the
higher--order contributions from these additional terms will generate further
cutoff dependence. The situation is familiar in field theory, where it is
recognized that renormalization requires all possible interactions allowed
by the symmetries. (For work in this direction, see Ref. \cite{doba}.) 
In this case, to have any predictive power, one should 
be able to argue that some ``sub--leading'' terms should be included in 
first order only when ``leading'' terms are included in second order.
For $\alpha=1$, this could be the case for the four--derivative two--body,
two--derivative three--body, and no-derivative four--body terms.

As we have shown here, the requirement of renormalizability
constrains the form of the interactions allowed at different orders in the
expansion beyond mean field. It calls
for a more general study where a
systematic analysis of the correct power counting within the perturbative
many--body problem with effective interactions is performed. To our
knowledge, this aspect, which we reserve for future work, has not been
addressed so far in the framework of the energy--density functional theories
based on Skyrme interactions. Once this is done, renormalizability could be
investigated in the context of potentially better--grounded interactions,
such as those that include pion effects in addition to the most general
short--range interactions (see, \textit{e.g.}, Ref. \cite{chiralmodel} and
references therein).

\medskip
\section*{Acknowledgments}

This research was supported in part by the 
by the U.S. Department of Energy, Office of Science, 
Office of Nuclear Physics, under award number DE-FG02-04ER41338, 
and by the European Union Research and Innovation program Horizon 2020 
under grant agreement no. 654002. 

\medskip
\section*{Appendix}

Defining the combinations of Skyrme parameters 
\begin{eqnarray*}
&&d_0=t_0^2\left(1+x_0^2\right)\ge 0, 
\\
&&d_1=2t_0\left[1+\frac{3}{16}\alpha(\alpha+3)+x_0x_3\right], 
\\\
&&d_2=\left[1+\frac{3}{16}\alpha(\alpha+3)\right]^2+x_3^2>0, 
\\
&&e_0=t_0 t_1(1+x_1 x_0), 
\\
&&e_1=t_1\left[1+\frac{3}{16}\alpha(\alpha+3)+x_1 x_3\right], 
\\
&&h_1=t_1^2\left(1+x_1^2\right)\ge 0, 
\\ 
&&h_2=t_2^2\left[1+x_2^2+4\left(1+x_2\right)^2\right]\ge 0,
\end{eqnarray*}
we can write the coefficients $A_i$, $B_i(\tilde{\lambda})$, and $C_i$
appearing in Eqs. \eqref{terms1}, \eqref{terms2}, and \eqref{terms3},
respectively, as 
\begin{eqnarray*}
&&A_0 = d_0 I_1, 
\quad 
A_1 = d_1 I_1, 
\quad 
A_2 = d_2 I_1,
\\
&&A_3 = e_0 I_2, 
\quad
A_4 = e_1 I_2, 
\quad
A_5 = h_1 I_3 + h_2 I_4, 
\\
&& B_0(\tilde{\lambda}) = d_0 + \frac{\tilde{\lambda}^2}{3}e_0 
+ \frac{\tilde{\lambda}^4}{20}h_1, 
\\
&& B_1(\tilde{\lambda}) = d_1 + \frac{\tilde{\lambda}^2}{3}e_1, 
\\
&&B_2(\tilde{\lambda}) = \frac{3}{5} e_0 
+ \frac{\tilde{\lambda}^2}{90}\left(\frac{27}{4}h_1 +h_2\right), 
\\
&&C_0 = d_2,
\quad
C_1=\frac{3}{5}e_1, 
\quad 
C_2=\frac{1}{70} \left(9 h_1 + h_2\right),
\end{eqnarray*}
where 
\begin{eqnarray*}
I_1=\frac{11-2\ln 2}{140}, &\quad& I_2=\frac{167-24 \ln 2}{1890}, \\
I_3=\frac{4943-564 \ln 2}{166320}, &\quad& I_4=\frac{1033-156 \ln 2}{498960}.
\end{eqnarray*}

\end{document}